\documentclass[pdflatex,sn-mathphys-num, iicol]{sn-jnl}

\usepackage{graphicx}%
\usepackage{multirow}%
\usepackage{amsmath,amssymb,amsfonts}%
\usepackage{amsthm}%
\usepackage{mathrsfs}%
\usepackage[title]{appendix}%
\usepackage{xcolor}%
\usepackage{textcomp}%
\usepackage{manyfoot}%
\usepackage{booktabs}%
\usepackage{algorithm}%
\usepackage{algorithmicx}%
\usepackage{algpseudocode}%
\usepackage{listings}%

\usepackage{todonotes}
\usepackage{xspace}
\usepackage{lineno}
\usepackage{rotating}
\usepackage{subcaption}
\usepackage{lineno}

\usepackage{makecell}

\newcommand {\ie}{\mbox{i.e.}\xspace}

\newcommand{\ET}{\ensuremath{E_{\mathrm{T}}}\xspace}

\newcommand{\HH}{HH\xspace}

\newcommand{\MADGRAPH} {\textsc{MadGraph}\xspace}
\newcommand{\MCATNLO} {\textsc{mc@nlo}\xspace}
\newcommand{\MGvATNLO}{\MADGRAPH{}5\_a\MCATNLO}
\newcommand{\PYTHIA} {{\textsc{pythia}}\xspace}
\newcommand{\PYTHIAEIGHT} {{$\PYTHIA\,8$}\xspace}

\newcommand{\DELPHES} {\textsc{Delphes}\xspace}

\newcommand{\II} {\textsc{II}\xspace}
\newcommand{\HLSML} {\textsc{HLS4ML}\xspace}
\newcommand{\ns} {ns\xspace}
\newcommand{\us} {\ensuremath{\mu{\mathrm s}}\xspace}
\newcommand{\kHz} {kHz\xspace}
\newcommand{\MHz} {MHz\xspace}

\theoremstyle{thmstyleone}%
\theoremstyle{thmstyletwo}%

\theoremstyle{thmstylethree}%

\begin{document}

\title[Article Title]{FPGA Implementation of a CNN-based Topological Trigger for HL-LHC}


\author*[1]{\fnm{J.} \sur{Brooke}}\email{james.brooke@bristol.ac.uk}

\author[1]{\fnm{E.} \sur{Clement}}\email{emyr.clement@bristol.ac.uk}
\author[2]{\fnm{M.} \sur{Glowacki}}\email{maciej.mikolaj.glowacki@cern.ch}
\author[1]{\fnm{S.} \sur{Paramesvaran}}\email{sudarshan.paramesvaran@bristol.ac.uk}
\author[1]{\fnm{J.} \sur{Segal}}\email{jeronimo.segal@bristol.ac.uk}

\affil[1]{\orgdiv{H H Wills Physics Laboratory}, \orgname{University of Bristol}, \orgaddress{\street{Tyndall Avenue}, \city{Bristol}, \postcode{BS8 1TL}, \country{United Kingdom}}}

\affil[2]{\orgname{CERN}, \orgaddress{\street{Esplanade des Particules 1}, \city{1211 Geneva}, \postcode{23}, \country{Switzerland}}}


\abstract{The implementation of convolutional neural networks in programmable logic, for applications in fast online event selection at hadron colliders, is studied. In particular, an approach based on full event images for classification is studied, including hardware-aware optimisation of the network architecture, and evaluation of physics performance using simulated data. A range of network models are identified that can be implemented within resources of current FPGAs, as well as the stringent latency requirements of HL-LHC trigger systems. A candidate model that can be implemented in the CMS L1 trigger for HL-LHC is shown to be capable of excellent signal/background discrimination for a key HL-LHC channel, HH(bbbb), although the performance depends strongly on the degree of pile-up mitigation prior to image generation.}




\maketitle

\section{Introduction}
\label{sec:intro}
Online event selection is a major experimental challenge at the Large Hadron Collider (LHC) \cite{Evans_2008}. The LHC general purpose experiments, ATLAS~\cite{ATLAS_2008} and CMS~\cite{CMS_2008}, produce data at exceptionally high rates, due to the high granularity of the detectors, with 4$\pi$ coverage, and the 40~MHz collision rate. In order to reduce the data rate to a level that can be stored for offline analysis, these experiments require complex ``trigger'' systems that can identify and select collision events of interest. In both experiments, the first stage of the trigger utilises a system of custom electronics using Field Programmable Gate Array (FPGA) technology, communicating via high bandwidth optical links \cite{ATLAS:2013lic, Tapper:1556311} that processes reduced granularity data, to generate a decision to accept/reject each event within a few microseconds. Subsequent stages of the trigger system rely on software processing to further reduce the data rate, before selected events are stored permanently for offline analysis.
The physics performance of these trigger systems, in terms of efficiency to select signals of interest, and the trigger rate, must be exceptional, given the challenge of searching for low cross-section physics at increasingly high instantaneous luminosity and large numbers of simultaneous proton-proton interactions, known as pile-up (PU). Typically, the logic implemented in the first level trigger performs a crude reconstruction of the collision, starting with detector level information and building up to physics objects such as leptons and jets, which are then used to select events by placing requirements on transverse energy ($E_T$), and other kinematic and topological criteria. The current L1 trigger systems at ATLAS and CMS must generate a decision within a latency of 2.5~\us and 3.8~\us, respectively, and with a maximum accept rate of ${\sim}~100~{\rm kHz}$. 

These requirements will be evolving as the next phase of the LHC begins in 2030. During LHC Run 1-3 the typical mean PU has been steadily increasing and is currently at the level of 55-60. The High Luminosity LHC (HL-LHC)~\cite{ZurbanoFernandez:2020cco} will deliver an order of magnitude more integrated luminosity to the general purpose experiments at an average PU of 200. In order to prepare for this unprecedented amount of data, experiments must refine and upgrade their trigger systems. The maximum accept rate of the ATLAS (CMS) triggers will increase to 1~\MHz (750~\kHz) and the available latency will increase to 10~\us (12.5~\us), but these measures alone are insufficient to maintain physics performance~\cite{Tapper:2013yva, CERN-LHCC-2017-020}.  Hence more sophisticated techniques, including additional detector inputs and more advanced algorithms, are being explored.
Furthermore, while approaches based on commercial hardware and software-based triggers are being pursued, the challenges foreseen at the Future Circular Collider (FCC-hh) mandate a low latency hardware trigger \cite{FCC:2018vvp}. Low latency implementations of advanced data processing algorithms may therefore have a significant impact on the physics potential of future colliders.

In terms of data processing algorithms for online event selection, the use of machine learning (ML) is a highly active area. Machine learning has been ubiquitous in offline analysis at collider experiments for many years, with increasing adoption in software-based online trigger systems~\cite{Albertsson:2018, Guest:2018}. Recent work has included firmware-based implementations of ML algorithms, which facilitate applications in ultra-low latency triggers. The implemented algorithms include neural networks~\cite{Duarte_2018,Nottbeck:2019rqu,Aarrestad:2021zos, Iiyama:2020wap, Bortolato:2024uqg, BurazinMisura:2024cuu, Alimena:2020web, Migliorini:2021fuj, Ospanov:2022fke, CMS-DP-2022-021, Aad:2021tru, Sun:2023, Aad_2023}, autoencoders ~\cite{Govorkova:2021utb, Bhattacherjee:2023evs,MeyerzuTheenhausen:2022ffb}, boosted-decision trees~\cite{Summers:2020xiy} and transformers~\cite{Jiang:2024lvg}. 

In this paper, we explore the use of simple computer vision methods to classify proton-proton interactions for use in hardware trigger systems at the HL-LHC. We transform detector data into 2D images, which are then processed by convolutional neural networks (CNNs). CNNs can achieve good image classification performance, using limited computational resources, due to two characteristics. First, the convolution operation exploits translational invariance by utilising shared kernels across the whole image, enabling weights to be derived from all locations on the input. Secondly, CNNs learn spatial hierarchies of patterns.  A first convolutional layer will learn small local patterns such as edges, a second convolutional layer will learn larger patterns made of the features of the first layers, and so forth. This allows CNNs to efficiently learn increasingly complex and abstract visual concepts.

This paper is arranged as follows; in Section \ref{sec:image} we discuss how the images could be generated within the HL-LHC L1 trigger systems, then in Section \ref{sec:simulation} we detail how we simulated such images for this study. Section \ref{sec:network} provides an overview the CNN model architecture, and Section \ref{sec:hwOpt} discusses hardware specific optimisations of the network.  In Section~\ref{sec:physics} we present the physics performance of this approach and in Section~\ref{sec:conclusions} we summarise the key findings of this study.

\section{Image Generation in HL-LHC Triggers}
\label{sec:image}
\begin{figure*}[hbtp!]
    \centering
    \includegraphics[width=0.8\textwidth]{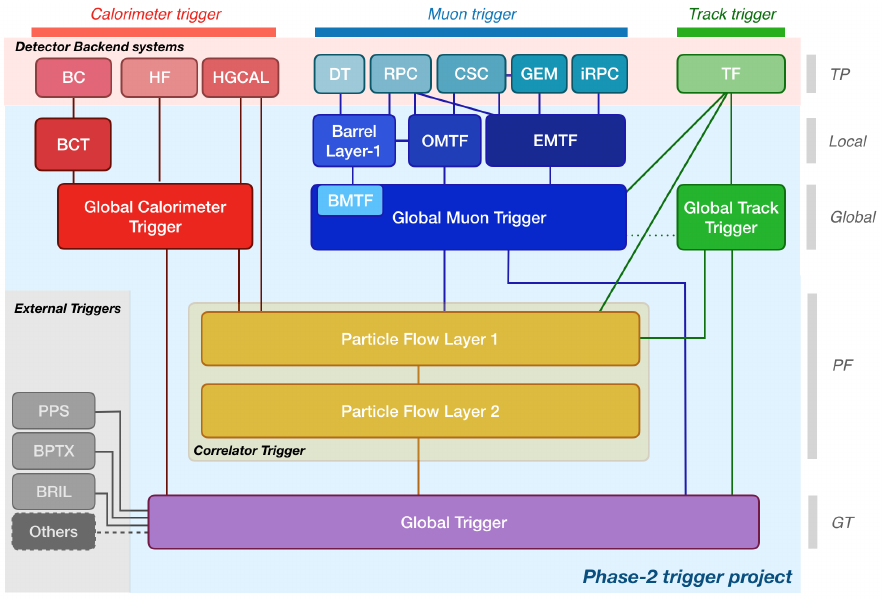}
    \caption{The CMS L1 trigger architecture for HL-LHC. Reproduced from \cite{Tapper:2013yva}, see therein for a full description of the components shown.
    }
    \label{fig:cmsp2trig}
\end{figure*}

In preparation for the HL-LHC, both ATLAS and CMS will upgrade their trigger and data-acquisition systems.  A key driver for these upgrades is the need to maintain trigger performance under increasing numbers of simultaneous proton collisions. Extracting signal information in this environment is highly challenging. These upgrades include use of detector information not currently available in the hardware trigger - for example, inclusion of particle trajectory data close to the interaction point (tracking information) at the full LHC bunch crossing rate of 40~\MHz in the CMS L1 trigger. They will also facilitate the use of larger FPGAs, and the latest FPGA technology, for example, AMD Versal Premium is targetted by the ATLAS L0 global processor ~\cite{Qian:2024cma}. Inter-board communication will use link speeds of up to 25~Gbps. Finally, advanced data processing architectures such as time-multiplexing are planned for both CMS and ATLAS.  This architecture segments the detector inputs to the trigger systems by bunch crossing, sending information from consecutive crossings to consecutive nodes in a round-robin fashion. This allows algorithms to be implemented that rely on data from the full detector, as well as accommodating algorithms requiring relatively long latency. As an example, the CMS L1 trigger architecture for HL-LHC is shown in Figure~\ref{fig:cmsp2trig}.

The hardware trigger architecture for both CMS and ATLAS at the HL-LHC provide several opportunities for representation of events as a 2D image. Perhaps most obviously, data from the calorimeters is typically segmented into small regions which are then sent to the hardware trigger. These are called ``calorimeter towers'' and can be visualised using the dimensions pseudorapidity $\eta$ and azimuthal angle $\phi$. These are commonly used in collider experiments; $\eta$ is a function of the angle with respect to the beam axis, and $\phi$ is an angle in the plane orthogonal to the beam axis. This data is easily represented as a 2D image for each bunch crossing, with pixel coordinates indicating $\eta$ and $\phi$, and pixel brightness representing $E_T$. Such images are easily and quickly generated, but will include energy from all pile-up events at nominal HL-LHC conditions.  Images with significantly reduced pile-up can be generated in the CMS Correlator Trigger (CT). This subsystem can be seen in Figure~\ref{fig:cmsp2trig} represented by the yellow boxes; here, data from the calorimeter, muon and tracking detectors are brought together, allowing a fast version of the particle flow algorithm to be applied, followed by the pile-up per particle (PUPPI) pile-up rejection algorithm~\cite{CMS:2017yfk,CMS:2020ebo,Tapper:2013yva}.  The resulting particle flow candidates aim to reconstruct individual particles, and can be histogrammed to give a 2D image representation of an event. Again, pixel brightness will represent the transverse energy, but in this case the pixels do not need to align with the calorimeter towers.

\section{Simulated Images}
\label{sec:simulation}

To quantify the potential performance of ML image classification in these contexts, we generated several image datasets that represent example signal and background images produced at the HL-LHC. Signal events were produced by simulating di-Higgs production, with both Higgs bosons decaying to a pair of b quarks. The matrix element calculations were performed with \MGvATNLO v2.9.9~\cite{Alwall:2014hca,Frederix:2018nkq}, at NLO accuracy in QCD. \PYTHIAEIGHT v8.305~\cite{bierlich2022comprehensive} was then used for the simulation of the parton shower, hadronization, and underlying event. Background events were produced by simulating minimum bias events, where a proton-proton interaction results only in soft multiple-parton interactions, using \PYTHIAEIGHT. The detector response was then simulated using \DELPHES v3.5.0~\cite{de_Favereau_2014,Selvaggi:2014mya,Mertens:2015kba}, which provides a parameterised description of the response of the CMS detector.

The impact of pile-up was simulated in \DELPHES by overlaying each event with a fixed number of 200 minimum bias events (200 PU).  Background events were generated, comprising only the overlaid pile-up interactions, \ie without the inclusion of the signal \HH process, corresponding to a typical bunch crossing in the CMS detector.

Images were then generated from the simulated calorimeter towers, which in CMS correspond to the sum of energy deposited in a group of 5x5 electromagnetic calorimeter (ECAL) cells and the hadronic calorimeter cell that is aligned with the ECAL cells. Images were generated by forming a two-dimensional (2D) histogram in $\eta$-$\phi$ of the transverse energy (\ET) deposited in each calorimeter tower. Only the region $|\eta| < 3$ was included, since PU rejection is expected to be most effective within this region, due to acceptance of the tracker and high-granularity endcap calorimeter detectors. The towers have a size of $0.083\times0.087$ in $\eta$-$\phi$, giving a default image size of $72 \eta \times 72 \phi$ pixels.  Images with lower resolution were also produced by downsampling these images.

A total of 64k signal events and 64k background events were generated to train the models. A further 16k of signal and 16k of background events were used as a validation sample during the training procedure. We verified that increasing the size of the training sample did not significantly improve the performance of either the model finally selected for implementation (which we will later label Model 3) or the largest model trained (Model 16). A further independent dataset comprising of 20k signal events and 200k background events were used for testing the physics performance of the models.  A significantly larger background sample was used in this case to accurately test background rejection of $\mathcal{O}(10^{4})$, which is comparable to that of the L1 trigger seeds discussed in Ref.~\cite{Tapper:2013yva}.

We also generated samples to explore the advantage of using pile-up subtracted data for image generation, such as may be obtained from the CMS CT.  We performed a crude pile-up rejection within the \DELPHES simulation by randomly rejecting, with fixed probability, particles that originate from pile-up interactions before they are included in the simulated calorimeter response. Three sets of signal and background samples were generated to simulate different PU mitigation conditions.  The first set was produced from the 200 PU samples, with a PU rejection probability of 70\%, representing a conservative estimate of the performance achievable by the CMS L1 PUPPI algorithm, based on information found in \cite{CMS-DP-2023-023}. The second set was produced with a PU rejection probability of 30\%, representing a scenario where PU rejection is much worse than expected. Finally, a sample was produced with no PU overlaid, to simulate the limit of perfect PU rejection. These samples do not attempt to simulate the PUPPI algorithm, rather they allow us to quantify the impact of pile-up rejection on image classification performance.

\begin{figure*}[hbtp!]
    \centering
    \begin{subfigure}{0.45\textwidth}
        \includegraphics[width=\linewidth]{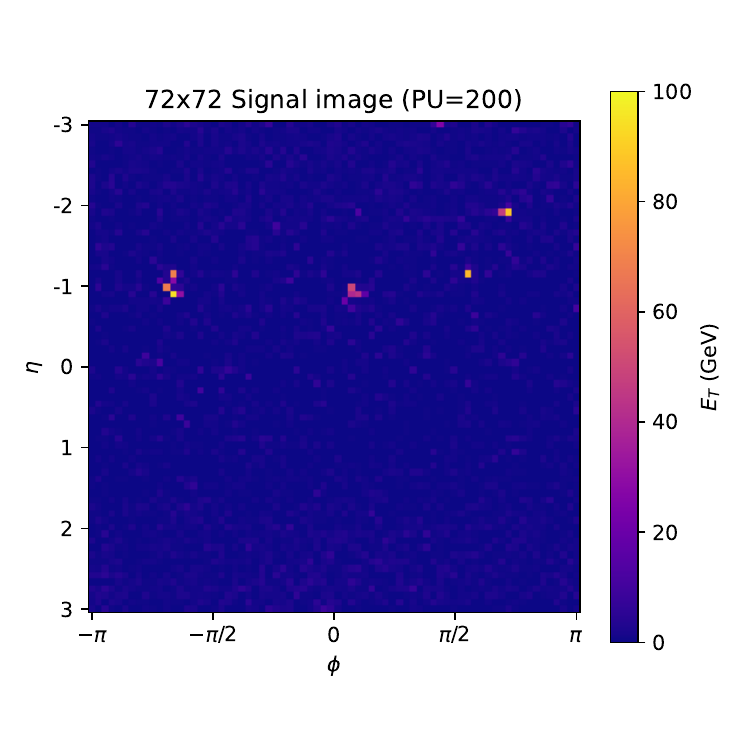}
        \subcaption[]{}\label{fig:example-images-a}
    \end{subfigure}
    \begin{subfigure}{0.45\textwidth}
        \includegraphics[width=\linewidth]{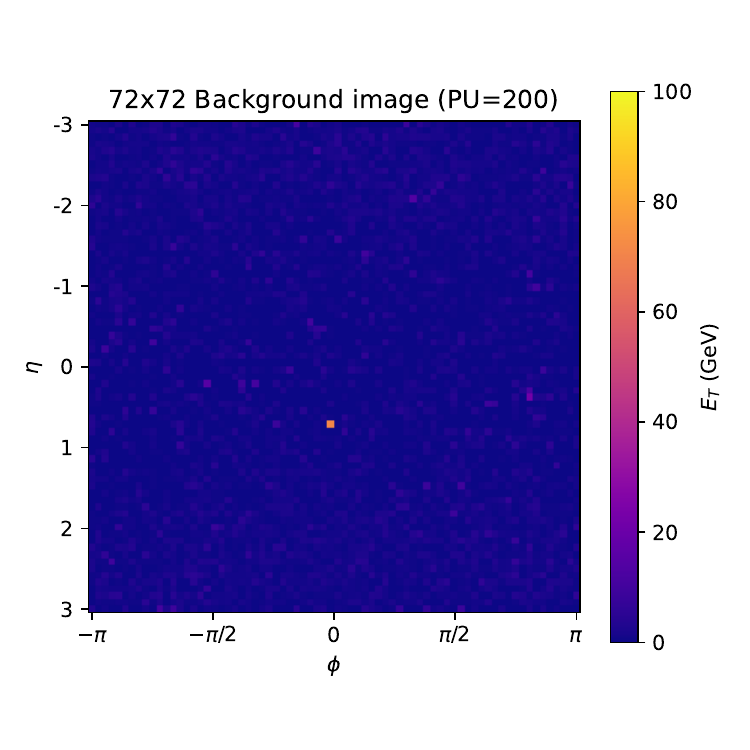}
        \subcaption[]{}\label{fig:example-images-b}
    \end{subfigure}
    \begin{subfigure}{0.45\textwidth}
        \includegraphics[width=\linewidth]{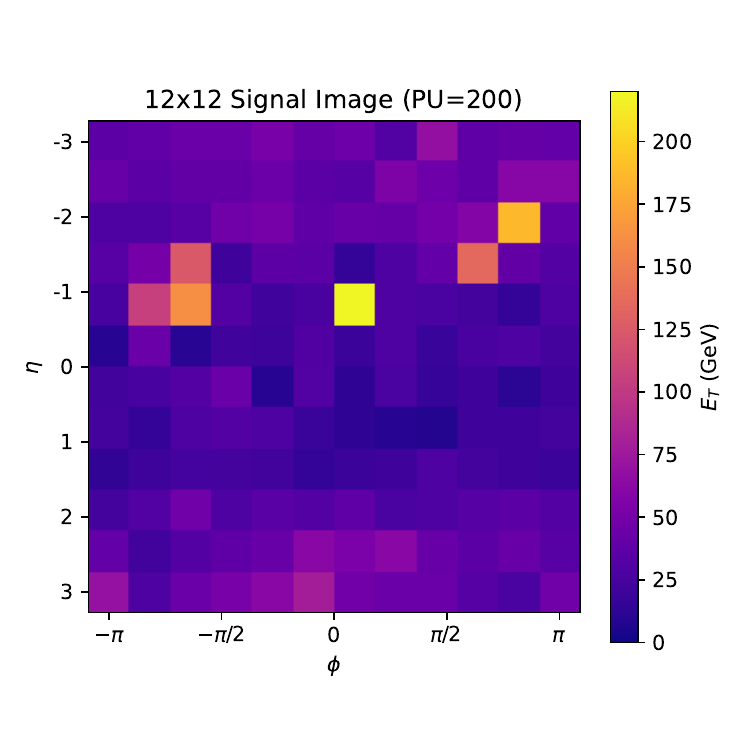}
        \subcaption[]{}\label{fig:example-images-c}
    \end{subfigure}
    \begin{subfigure}{0.45\textwidth}
        \includegraphics[width=\linewidth]{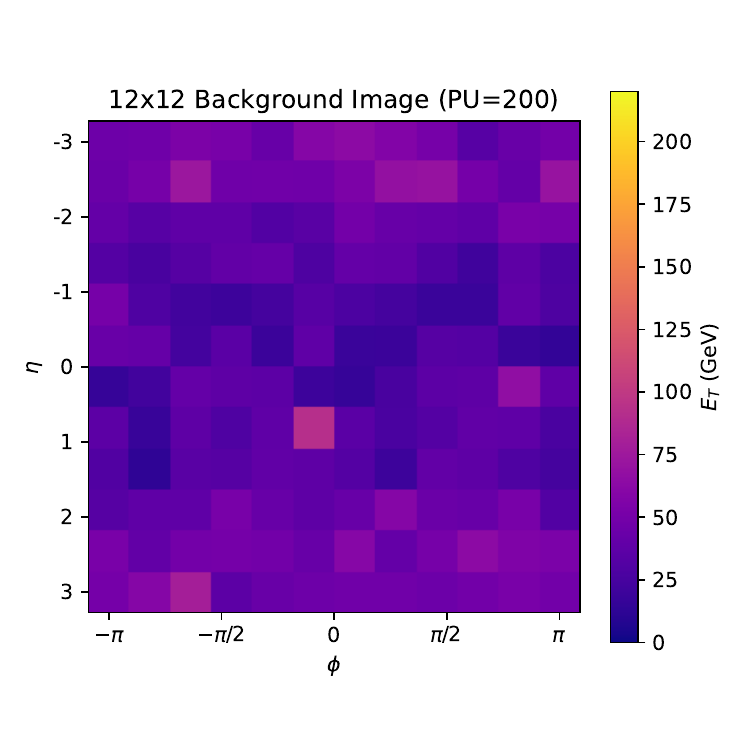}
        \subcaption[]{}\label{fig:example-images-d}
    \end{subfigure}
    \caption{Example images used for model training and evaluation, for 200 PU. A signal HH(bbbb) image is shown at a) $72\times72$ pixel resolution, and b) $12\times12$ pixel resolution. A background image is shown with c) $72\times72$ pixel resolution, and d) $12\times12$ pixel resolution. }
    \label{fig:example-images}
\end{figure*}

Example images are shown in Figure~\ref{fig:example-images}, for HH(bbbb) and minimum bias processes with 200 pile-up, for image sizes of $72\times72$ and $12\times12$ pixels.

\section{Network Architecture}
\label{sec:network}

\begin{figure*}[hbtp!]
    \centering
    \includegraphics[width=0.8\textwidth]{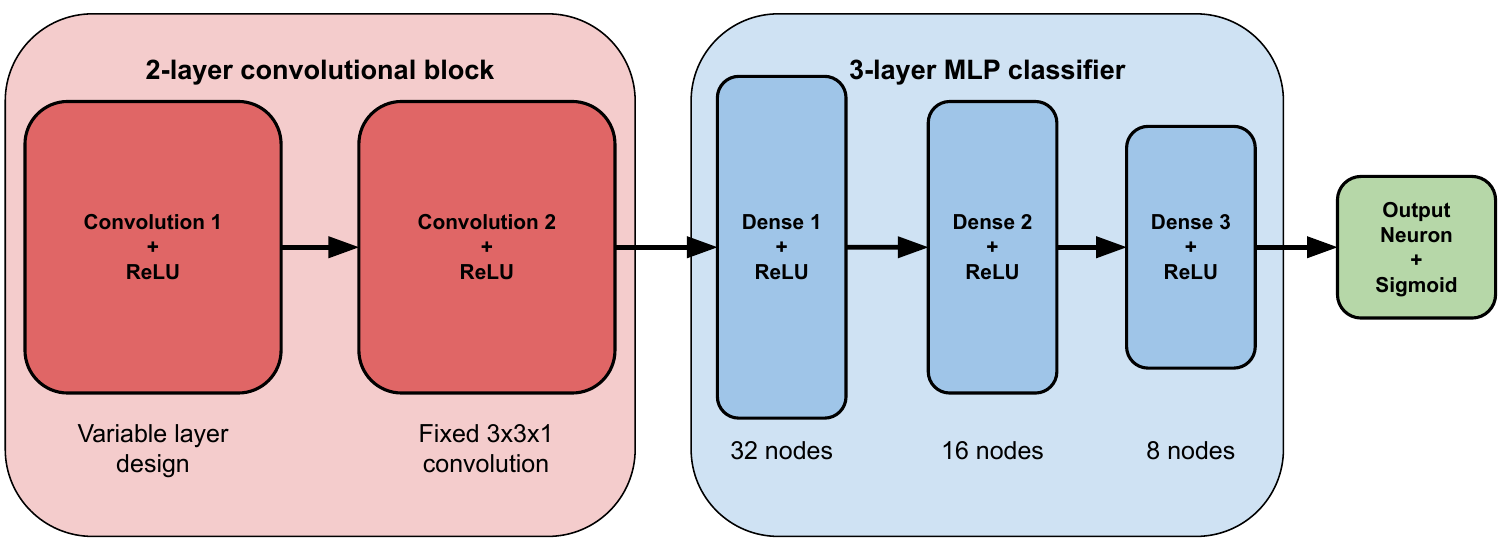}
    \caption{A schematic of the model architectures under study.  The convolutional layers are shown in red, and the fully connected layers are shown in blue.  The parameters of the fully connected layers and the second convolutional layer are the same for all model architectures, whereas the parameters of the first convolutional layer, and the size of the inputs, differ for each model.}
    \label{fig:network-diagram}
\end{figure*}

Our approach relies on classification of images as either signal or background using a CNN.  The CNN architectures studied here share a similar design, which is shown schematically in Figure~\ref{fig:network-diagram}, and are composed of two modules. The first module, tasked with feature extraction, consists of two convolutional layers, each followed by a ReLU activation function for non-linearity.  Following the feature mapping performed by the convolutional layers, the output tensor is flattened and passed to a fully connected network consisting of three layers with 32, 16 and 8 neurons respectively. Each fully connected layer is again followed by a non-linear ReLU activation function. Finally, a single neuron output layer is used for binary classification, with the output score being translated to a probability distribution by the sigmoid activation function.  The model architectures being studied differ in the size or granularity of the input images, the size of the kernel in the first convolutional layer, the stride of the kernel in the first convolutional layer, and the number of filters in the first convolutional layer.  The architecture of the second convolutional layer is the same for all models considered, and comprises one kernel of size three, with a stride of one.

When training image recognition networks, gradient descent convergence and numerical stability is greatly improved if all images have pixel values in the range [0,1]. Due to the wide dynamic range of the \ET of particles entering our detector simulation, and hence the pixels in our images, we first saturate images to a maximum \ET value, setting all \ET values above this to the maximum value.  The \ET values are then scaled to the [0,1] range. The value for saturating the pixel \ET was set to 512~GeV, which was found to maximise the performance of the models based on the fixed-rate efficiencies described in Section~\ref{sec:physics}.

Another consideration that has to be addressed is the discontinuity of unrolling cylindrical geometries to form 2D images; after this procedure the model is not intrinsically aware that opposing edges in the $\phi$ plane are connected. Events where physics level objects with a typical size of many pixels, such as jets, populate the edges of the images in the $\phi$ dimension can cause misclassification of the event. To circumvent this, input images are padded by translating three pixel columns from one side of the image to the opposing side.  This choice ensures that a jet that is centred near the edge of the image in $\phi$ is represented at least once within the padded image.  Finally, to allow the convolution kernel to be centred on edge pixels, the images are zero-padded in the $\eta$ plane.

The models were implemented and trained using the Keras API~\cite{chollet2015keras}.  The training was performed using the ADAM optimiser~\cite{kingma2017adammethodstochasticoptimization}, with the binary cross entropy as the loss function being minimised. As the models are targeting deployment on an FPGA, we employ quantisation-aware training to ensure that fixed-point arithmetic is taken into account during the training procedure.  The QKeras package~\cite{Coelho_2021} was used to implement the quantised models, and the weights of all layers except the final layer are quantised to a precision of 8 bits in total, with 2 integer bits.  The output node and the final sigmoid layer utilizes a greater number of bits, 16 in total with one integer bit, to maintain precision in the output score. This allows us to measure the efficiencies over a wide range of rates with fine granularity. The performance of the quantised model with this level of quantisation is found to have the same performance as a model utilising floating-point arithmetic.   To prevent overfitting, the training is halted when the loss function calculated on the validation set is no longer decreasing over a defined number of epochs.

The choice of quantisation places a significant restriction on the magnitude of the weights that can be learned. Weights were prevented from growing too large, and potentially exceeding the maximum representable value, by adding a term to the loss function that is proportional to the magnitude of the weight squared, i.e. L2 regularization~\cite{NIPS1988_1c9ac015,NIPS1990_bc6dc48b,NIPS1991_8eefcfdf}. A regularization strength of $10^{-5}$ was found to provide sufficient regularization without sacrificing model performance.



\section{Hardware Aware Model Optimisation}
\label{sec:hwOpt}

The network architectures described above have a large parameter space. Whilst our approach relies on automatic firmware generation, following this through to an implementation which meets hardware and timing constraints and is ready for deployment typically requires manual intervention. We therefore reduce the model space using several steps, first by identifying a shortlist of potentially implementable models by calculating the total number of operations, then by generating HDL for these models and estimating resource usage with Vivado HLS. Finally, we examine the physics performance of these models, and select a single model which is synthesised and implemented using Vivado, including manual tuning to ensure clock constraints are achieved.

In order to assess whether FPGA implementation of a network is feasible, we must assume a particular application.  For this study, we assume images are generated in the CMS CT, where the target FPGA is the AMD VU9P~\cite{uscaleplus}. We require the network to be implementable on a single Super Logic Region (SLR) of this FPGA, corresponding to approximately one third of the total logic available.  We choose an FPGA clock frequency of 360~\MHz and we assume the CT has a time-multiplexing period of 6 bunch-crossings. This results in a maximum initiation interval (\II) of 54 clock cycles, or 150~\ns, where \II is defined as the period of time after which the model can accept data from a new event.

To explore the model design space that can be accommodated in the target FPGA and meet the constraints set by the chosen application, we scan a range of networks with different parameters and consider those that meet the criteria on the \II and resource usage. The primary parameter varied for the scan of potential networks was the input image resolution. Higher-resolution images offer more detailed information about the event, but also quadratically increase the computational cost of the first convolutional layer, and the cost of achieving a desired \II. In the following, we consider images with a resolution of $72\times72$, and downsampled images with resolutions of $36\times36$, $24\times24$, $18\times18$, and $12\times12$ pixels.

For each input image size, we adjusted the kernel size and stride in the first convolutional layer to ensure all pixels are seen by the model.  Increasing the kernel size and stride will reduce the firmware resources of a model by reducing the number of instances of filters that are computed in a convolutional layer, which also reduces the \II of the model.  The number of filters in the first convolutional layer was varied when resource estimates allowed.  

The hyperparameters of the 2nd convolutional layer and the dense layers were optimised using a random search with KerasTuner~\cite{omalley2019kerastuner} using a model based on images with a resolution of $18\times18$, and a single kernel size of 5 and stride of 1 in the first convolutional layer.  Since the choices of kernel size and stride of the first convolutional layer for each model reduce the dimensions of the input images to a similar size for all models, we use the same architecture choice for subsequent layers in all models.

For each model in the scan over image size and configuration of the first convolutional layer, we compute the total number of operations required by the network.  We found that requiring this to be less than 8000 results in models that were likely to meet hardware constraints. Whilst this selection will produce a set of varied models, it should be noted that some selected models will fail to meet the hardware constraints once the models are synthesised by Vivado HLS. For the same reason, we will not necessarily obtain an exhaustive list of models capable of meeting the final hardware constraints. This requirement resulted in a shortlist of 16 models, which are listed in Table~\ref{tab:resource-use}.


For each model in Table~\ref{tab:resource-use}, the \HLSML package~\cite{fastml_hls4ml, Duarte_2018,Aarrestad:2021zos,Ghielmetti:2022ndm} was used to convert the model architecture and the corresponding weights into Vivado HLS code, which in turn was synthesised into VHDL code. An implementation of {\it im2col}~\cite{Aarrestad:2021zos,vasudevan2017parallelmultichannelconvolution} within HLS4ML was utilised, allowing multiple convolution operations to be performed in parallel.  This reduces the latency and the \II of the model when compared to a firmware implementation that accepts a single bin as an input on each clock cycle, at a cost of increased firmware resource usage.  The level of parallelization was chosen such that one output row is computed per clock cycle.  It may be possible to further reduce the resource usage of a particular model by fine-tuning this parameter, but since the first convolutional layer only contributes around ~10\% of LUT usage within an SLR, we have not explored this further. Similarly, pruning the models during training could reduce their LUT usage, but the \II would remain unchanged.  As several architectures are already found to satisfy the firmware constraints without pruning, we did not consider the effect of pruning on the resource usage in this study. 



\begin{figure}[hbtp!]
    \centering
    \includegraphics[width=0.45\textwidth]{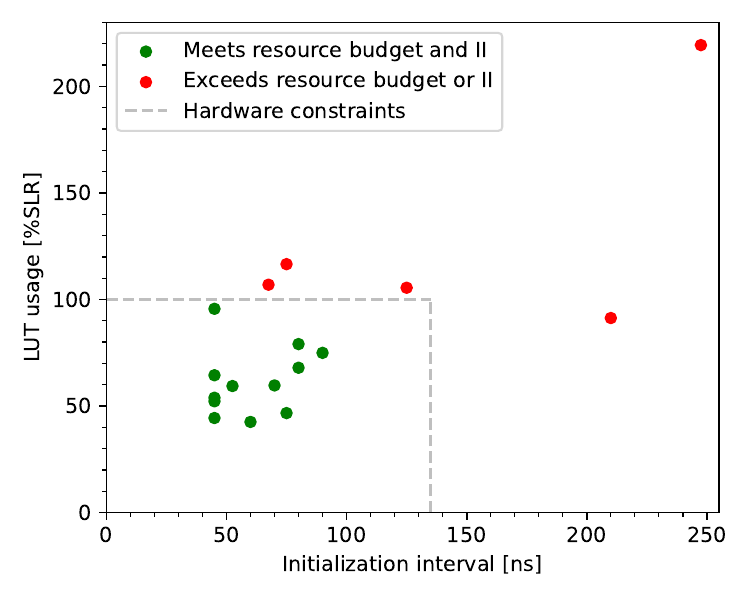}
    \caption{Estimated LUT usage against II for the initial scan over model architectures. Resource usage is estimated by Vivado HLS. The grey dashed lines indicate maximum values of LUT usage and II which can be accommodated within the target FPGA and application.}
    \label{fig:LUT-vs-II}
\end{figure}

Preliminary estimates of the resource usage of each model were obtained from Vivado HLS.  We observe the look-up table (LUT) usage to be the determining factor in whether a model fits within resource constraints, so these values are given in Table~\ref{tab:resource-use} along with the latency and \II, also obtained from Vivado HLS.  The LUT usage and \II are shown in Figure~\ref{fig:LUT-vs-II}, with models that meet the constraints set out above, shown in green. The flip-flop (FF), DSP and Block RAM (BRAM) usage estimates for these models are all small relative to the resources available, with FF usage ranging from 4\% to 9\% of an SLR, DSP usage 0\%-0.5\%, and BRAM usage up to 27\%.  The low DSP usage is expected given the quantisation of our models.  The latency of these models ranges from 228~\ns to 397~\ns, with models using fewer LUTs typically having a lower latency, and vice versa.  During the model selection, we did not include a constraint on the total latency of our models.  However, we note that the latencies we obtained of these 11 models are lower than those of other algorithms being developed for the CMS CT, such as a jet finder~\cite{CMS-DP-2023-023}, where a constraint of 1~\us was considered for the latency.  Whilst we have not considered the implications on the latency and resource usage of any preprocessing of the input data that may be required, we do not expect these to affect the firmware viability of the 11 models.
 
We note that the \II decreases from 24 clock cycles for Model 1, to  18 clock cycles for both Models 2 and 3.  These three models are based on the same image size, kernel size and stride in the first convolutional layer, and are expected to have the same \II, even though the number of filters in this layer differ.  This indicates that a different choice was made by the software tools during the synthesis of Model 1 into VHDL, compared to the other two models. 
 
A few models shown in red in Figure~\ref{fig:LUT-vs-II}, and also listed in Table~\ref{tab:resource-use}, do not meet the hardware constraints.  Models 12-14 satisfy the criterion on the \II, but the LUT usage is greater than the requirement.  Whilst the LUT usage could be reduced by pruning, or by other model compression techniques, these models are not considered further in this study, given that at least one other model based on the same image sizes, namely $18\times18$, $24\times24$, and $30\times30$, does satisfy the hardware constraints.  Model 15 is based on the lowest resolution images we consider, and is an attempt at constructing a model that meets the hardware constraints where the kernels in the first convolutional layer have a stride of one.  For this model, it was necessary to include an additional max pooling layer between the first and second convolutional layers to reduce the number of parameters to a level comparable with models 1-11.  Whilst this additional step results in a model that meets the criterion on the resource usage, the \II of this model is too large.  Model 16 has the largest II and resource usage, and is based on the maximum image size.  This model uses a single filter with a relatively large kernel size and the largest possible stride for this particular architecture.  This represents a minimal model that is capable of seeing every pixel in the maximum sized images.  Whilst it was not expected for such a model to satisfy the hardware constraints, this model will be used in Section~\ref{sec:physics} when considering the physics performance of our models.

\begin{table*}[t]
\centering
\caption{Key hyperparameters for the initial shortlist of 16 model architectures, including image size, kernel size, stride and number of filters in the first convolutional layer, and the total number of trainable parameters. Preliminary estimates of resource usage, \II and latency are also given, obtained from \HLSML and Vivado HLS, with AMD VU9P as the target FPGA. The LUT usage is quoted with respect to the total within an SLR, since this is the determining factor in whether a model can be implemented or not. FF and BRAM usage are omitted for clarity, since these values are small and not close to hardware constraints.  Models 1-11 satisfy the hardware constraints described in the text, whereas Models 12-16 do not.}\label{tab:resource-use}
\begin{tabular}{|c|c|c|c|c|c|c|c|c|c|}
\hline
\textbf{Model} & \textbf{Image size} & \textbf{Kernel} & \textbf{Stride} & \textbf{Filters} & \textbf{Params.} & \textbf{LUT} & \textbf{II} & \textbf{Latency} \\
\textbf{label} & \textbf{(padded)} & \textbf{size} & & & & \textbf{(\% of SLR)} & \textbf{(clock cycles)} & \textbf{(ns)} \\
\hline
1 & 18 & 3 & 3 & 1 & 1237 & 42.5 & 24 & 228 \\ 
2 & 18 & 3 & 3 & 2 & 1256 & 44.4 & 18 & 244 \\ 
3 & 18 & 3 & 3 & 4 & 1294 & 53.9 & 18 & 283 \\ 
4 & 18 & 4 & 2 & 1 & 1884 & 75.0 & 36 & 356 \\ 
5 & 18 & 6 & 2 & 1 & 1552 & 59.4 & 21 & 317 \\ 
6 & 24 & 3 & 3 & 1 & 1877 & 68.0 & 32 & 311 \\ 
7 & 24 & 4 & 4 & 2 & 1270 & 46.7 & 30 & 292 \\ 
8 & 24 & 6 & 3 & 1 & 1552 & 59.7 & 28 & 325 \\ 
9 & 30 & 5 & 5 & 1 & 1253 & 52.2 & 18 & 283 \\ 
10 & 42 & 7 & 5 & 1 & 1917 & 79.1 & 32 & 397 \\ 
11 & 42 & 7 & 7 & 1 & 1277 & 64.5 & 18 & 303 \\ 
\hline
12 & 24 & 6 & 2 & 1 & 2800 & 116.6 & 30 & 311 \\ 
13 & 24 & 8 & 2 & 1 & 2348 & 106.9 & 27 & 419 \\ 
14 & 30 & 3 & 3 & 1 & 2773 & 105.5 & 50 & 444 \\ 
15 & 18 & 5 & 1 & 1 & 1541 & 91.3 & 84 & 450 \\ 
16 & 78 & 8 & 7 & 1 & 3372 & 219.4 & 99 & 739 \\ 
\hline
\end{tabular}
\end{table*}

\section{Physics Performance}
\label{sec:physics}

Having determined a set of models which can be accommodated within  reasonable firmware resources, we train those models as described in Section~\ref{sec:network}.  To evaluate their performance, we assume events are selected by the online trigger system if the classification score exceeds a threshold. The L1 trigger rate is then computed as a function of threshold from the sample of minimum bias events, while the signal efficiency is computed in the same way from the sample of HH(bbbb) events.  The L1 trigger rate and signal efficiencies for our models are calculated at multiple points to construct a receiver operating characteristic (ROC) plot, and the area under the curve (AUC) is calculated to determine the performance of the models over a wide range of trigger rate and signal efficiency, covering several orders of magnitude of L1 trigger rate.  We also consider a single comparison point from the ROC curve that reflects a possible working point for such an algorithm.  In particular, we choose a point corresponding to the signal efficiency obtained for an L1 trigger rate of 10~\kHz, which is comparable to the rate of L1 trigger seeds discussed in Ref.~\cite{Tapper:2013yva} that would target HH(bbbb) events. Henceforth we refer to the efficiency at 10~kHz as the ``fixed-rate efficiency'' (FRE).


For Models 1-11, the AUC is found to be in the range $0.943\pm0.004$ to $0.982\pm0.002$, in the 200 PU samples without any PU mitigation.  The AUC is computed using $k$-fold cross-validation, with $k=5$, with the final AUC and its uncertainty obtained from the mean and standard deviation computed over the 5 folds.  Similarly, we obtain FRE values in the range $27\pm1$\% to $32\pm1$\%, in the 200 PU samples without any PU mitigation.  The greater relative uncertainty in the FRE values originates from the statistics in the background test sample used to determine the 10~kHz threshold, since this rate corresponds to a background rejection factor of $\mathcal{O}(10^{4})$.

We find Model 3 to have the highest mean AUC, as well as the highest AUC in 3 out of the 5 folds.  This model is based on $18\times18$ pixel images including padding, has a kernel size of three in the first convolutional layer, a corresponding stride of three, and a total of four filters.  This model is chosen as the final hardware-constrained model for the following discussion.  While we have used AUC here, it is worth noting that Model 3 also provides the highest mean FRE and the highest FRE in 3 out of 5 folds. 


Before further examining the physics performance of Model 3, we first check that the model continues to meet the hardware constraints after synthesising and implementing the model into firmware.  We found that LUT usage was reduced by 60\% following a full implementation, and the BRAM usage was halved.  However, the FF usage remained similar, and the DSP usage increased from 0.1\% to 4\%.  These differences imply some optimisation of logic has taken place during the Vivado synthesis and implementation, and in some cases a different choice of resource was made between the HLS synthesis and the firmware build.  In order to ensure the built firmware met signal timing requirements, additional registers needed to be inserted at several points in the data path, increasing the model latency by 118~\ns. Despite this increase, Model 3 continues to satisfy the initial hardware constraints, with additional margin on the LUT usage.

\begin{figure}[hbtp!]
    \centering
    \includegraphics[width=0.45\textwidth]{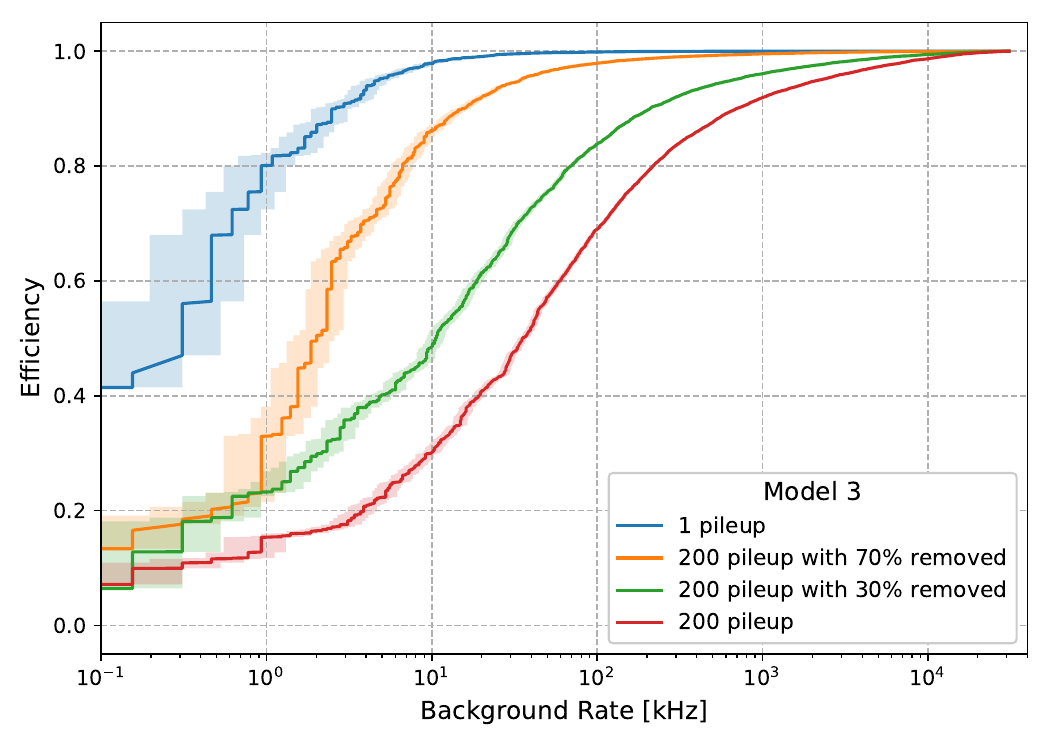} 
    \caption{Efficiency to select HH(bbbb) signal against background rate, for the hardware constrained model (Model 3), under 4 different pileup scenarios. The central values presented here are those for a single $k$-fold, and the shaded band indicates the uncertainty on the rate due to Monte Carlo statistics only. }
    \label{fig:model3-roc}
\end{figure}

Figure~\ref{fig:model3-roc} shows the HH(bbbb) signal efficiency against the background trigger rate for Model 3, under the different PU scenarios described in Section~\ref{sec:image}. As can be seen, in the absence of PU (i.e. the 1 PU scenario) the model is capable of achieving excellent signal efficiency (98\%) at modest trigger rate (10~kHz). This is consistent with the CNN learning key features of the signal from the full kinematic and topological information in the event, enabling excellent discrimination from the QCD background.  The presence of PU has a significant impact on the efficiency at this trigger rate, but the model can still achieve $\sim 85$\% ($\sim 50$\%) efficiency in the 200~PU scenario with 70\% (30\%) PU mitigation.

Existing studies for selecting HH(bbbb) at the Phase-2 CMS Level-1 Trigger using jets reconstructed from PUPPI candidates and exploiting the potential of b-tagging show signal efficiencies of 64\% at a trigger rate of 10~\kHz ~\cite{CMS-DP-2022-021}. We cannot directly compare our results to these studies, due to the differences between our naive PU mitigation and CMS L1T PUPPI.  We used the 70\% PU removal scenario as a conservative estimate of what can realistically be achieved.  If this is indeed the case, the CNN approach can be seen to compete with existing methods.  Given this, we recommend further studies of the CNN approach be performed using the full CMS detector and trigger simulation, including L1T PUPPI.

We now explore the effect of the hardware-based optimisation on the performance of our selected model. If the performance could be improved significantly beyond that of the hardware constrained model, further optimisation of the model may be useful. To study this, we compare Model 3 with a model based on the finest granularity images that comes close to meeting hardware constraints (Model 16), and a model based on finest granularity images that is not constrained. The last of these models uses a single $5 \times 5$ filter with a stride of 1 in the first convolutional layer, and comprises a total of 7677 parameters.  Based on the number of parameters, this model is at least a factor four larger than any of our hardware-constrained models, and over double the size of Model 16.  No attempt was made to synthesise the model into VHDL, however this would likely result in firmware that requires more resources than are available on the target FPGA, and a significantly larger \II than models shown in Table~\ref{tab:resource-use}. 

\begin{figure}[hbtp!]
    \centering
    \begin{subfigure}{0.45\textwidth}
        \includegraphics[width=\linewidth]{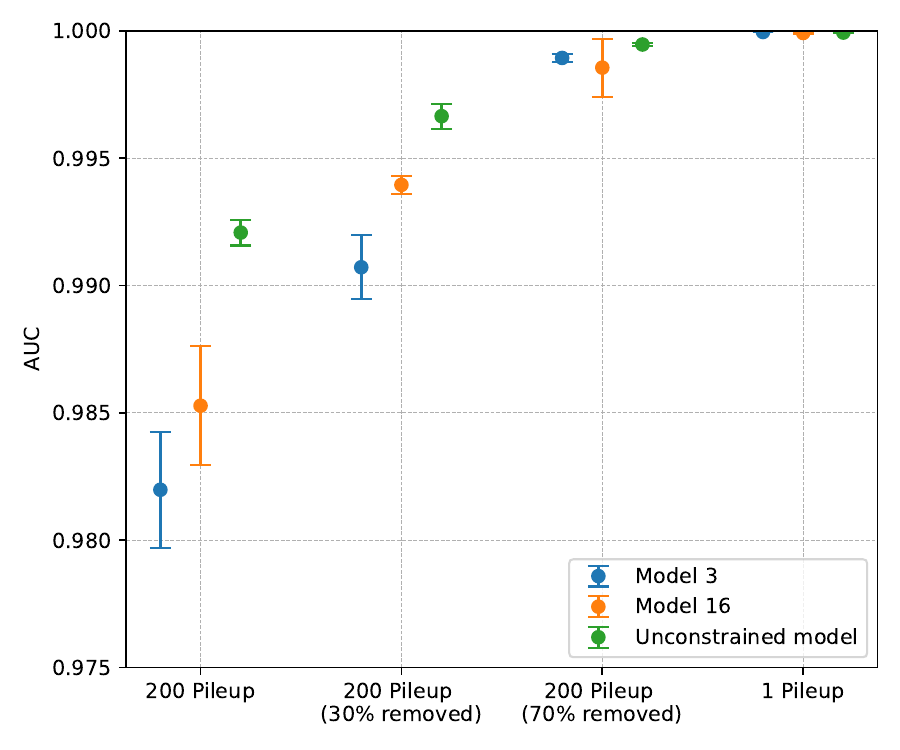}
        \subcaption[]{}\label{fig:auc-by-model}
    \end{subfigure}
    \begin{subfigure}{0.45\textwidth}
        \includegraphics[width=\linewidth]{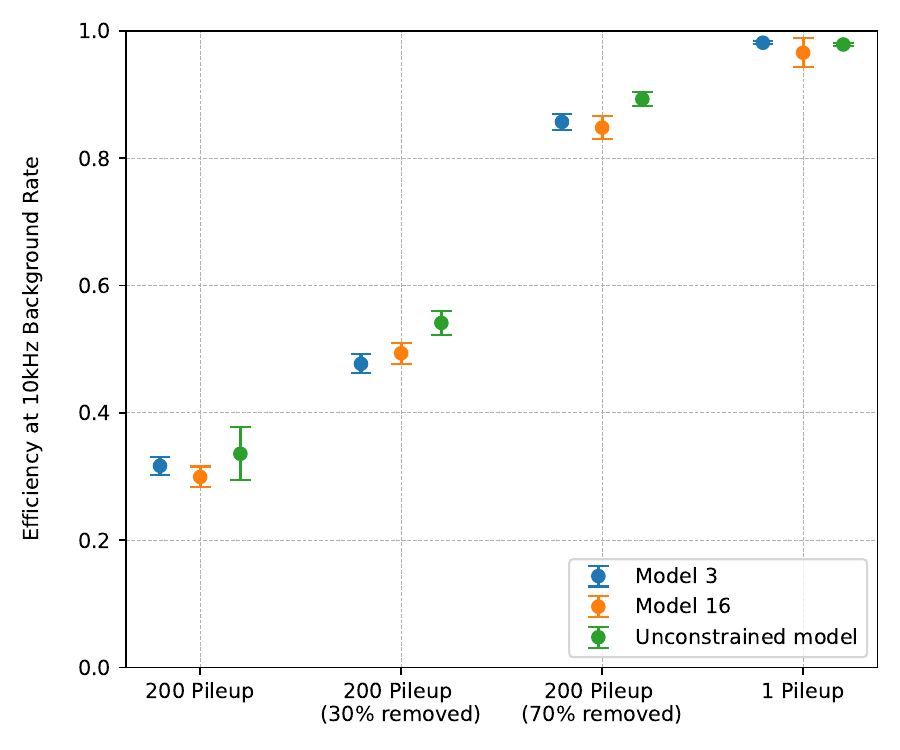}
        \subcaption[]{}\label{fig:fre-by-model}
    \end{subfigure}
    \caption{a) Area under the curve (AUC) and b) fixed rate efficiency (FRE) at 10~kHz, for Model 3, Model 16 and an Unconstrained model (see text for details).  The central values and uncertainty are obtained from the mean and standard deviation of each metric over 5 $k$-folds.}
    \label{fig:auc-fre-by-model}
\end{figure}

\begin{table*}[htbp!]
\centering
\caption{Area under the curve (AUC) and fixed rate efficiency (FRE) at 10~kHz, for Model 3, Model 16 and an Unconstrained model (see text for details).  The central values and uncertainty are obtained from the mean and standard deviation of each metric over 5 $k$-folds.}\label{tab:auc-by-model}
\begin{tabular}{|c|c|c|c|c|}
\hline
 & \textbf{200 PU} & \makecell{\textbf{200 PU} \\ \textbf{30\% removal}} & \makecell{\textbf{200 PU} \\ \textbf{ 70\% removal}} & \textbf{1 PU}  \\ \hline
 & \multicolumn{4}{c|}{\textbf{AUC}} \\  
\hline
Model 3 & $0.982 \pm 0.002$ & $0.991 \pm 0.001$ & $ 0.9989 \pm 0.0002$ & $0.999951 \pm 0.000002$\\ 
Model 16 & $0.985 \pm 0.002$ & $0.9940 \pm 0.0004$ & $0.999 \pm 0.001$ & $0.99991 \pm 0.00006$\\
Unconstrained & $0.9921 \pm 0.0005$ & $0.9966 \pm 0.0005$ & $0.99946 \pm 0.00007$ & $0.999928 \pm 0.000008$ \\ \hline
 & \multicolumn{4}{c|}{\textbf{FRE}} \\  
\hline
Model 3 & $32 \pm 1$\% & $48 \pm 1$\% & $ 86 \pm 1$\% & $98.1 \pm 0.2$\% \\ 
Model 16 & $30 \pm 2$\% & $49 \pm 2$\% & $85 \pm 2$\% & $97 \pm 2$\% \\
Unconstrained & $34 \pm 4$\% & $54 \pm 2$\% & $89 \pm 1$\% & $97.8 \pm 0.3$\% \\
\hline

\end{tabular}
\end{table*}

Figure~\ref{fig:auc-fre-by-model} shows the AUC and FRE obtained for each of these three models under the different PU scenarios described in Section~\ref{sec:image}. These values are also presented in Table~\ref{tab:auc-by-model}. As can be seen, the improvement in either AUC or FRE achieved by increasing the model size is much less than the difference between PU scenarios. However, we see that AUC broadly increases with model size in all PU scenarios.  For the 200PU (70\% removal) scenario this trend is less clear due to increased uncertainty on Model 16, however there is a marked improvement between Model 3 and the Unconstrained model.

The FRE results do not show the same clear trend as AUC with model size, apart from the Unconstrained model in the scenarios with PU removal, which has a few-\% improvement over the other two models. The uncertainty on FRE is dominated by the statistical uncertainty from the background test sample used to calculate rates, due to $\mathcal{O}(10^{4})$ rejection factor involved.  Increasing the size of this sample may reduce the uncertainty and reveal a correlation between model size and FRE. However, larger training samples may also be required, to allow the model to distinguish the discriminating features of background at large rejection factors. This was not pursued due to the prohibitively large computation time required, and because the results shown in Table~\ref{tab:auc-by-model} indicate that substantial gains from increasing model size seem unlikely.

We therefore conclude that, as well as using realistic simulation of PU mitigation, future optimisation of such a model might usefully employ more advanced training approaches, to focus the model on discriminating background features that are most important at 10~kHz rate.

\section{Conclusions}
\label{sec:conclusions}

In this study, we set out to assess the viability of Convolutional Neural Networks implemented in FPGAs, as a method of selecting events of interest at the High Luminosity LHC. We have shown that, provided the image size is sufficiently small, such networks can be implemented in FPGA technology planned for the CMS HL-LHC trigger, with the full network computation being handled within very tight latency constraints associated with the LHC trigger systems. In the absence of pile-up, a candidate network was shown to have excellent signal efficiency for a high value signature, HH(bbbb). If the input images can be generated after some moderate pile-up mitigation is applied, then the performance improves significantly over the no mitigation scenario. It should be noted, however, that these studies are limited by the use of fast simulation of the CMS detector, and an extremely simplistic pileup removal algorithm. Further studies using the full detector simulation and realistic pile-up mitigation algorithms will be required to fully evaluate the performance of this approach.

Since the approach studied here focuses on a single signal of interest, the bandwidth allocation and resource usage on an FPGA could be considered high, especially as conventional heuristic trigger paths typically use much lower resources per trigger line. However, we have explored a particularly high-value signal, that of di-Higgs production with each Higgs decaying to two bottom quarks. The physics benefits gained from improving selection of such events may outweigh the cost in terms of FPGA resource. Furthermore, as larger FPGAs and more ML-adept toolsets come to market, this should be less of an issue and as such the authors are currently investigating the use of the AMD Versal AI Core series, along with VitisAI~\cite{vitisai} and High-Granularity Quantisation~\cite{Sun:2024soe} to explore the benefits of this new technology, both for potential future upgrades of CMS and for future colliders such as FCC-hh. In these contexts it will become increasingly important to exploit the use of machine learning for ultra-low latency applications.

\backmatter





\bmhead{Acknowledgements}

We acknowledge the Fast Machine Learning collective as an open community of multi-domain experts and collaborators, which has made this work possible. We thank Dr Aaron Bundock for his feedback and help preparing this manuscript. This work was supported by the Science and Technology Facilities Council (grants ST/W000490/1, ST/R005788/1, ST/P006779/1), as well as the Engineering \& Physical Sciences Research Council (grant EP/S023992/1).

\bmhead{Author contributions}

All authors contributed to the research methodology, discussion of results, and review and editing of the manuscript. JB conceived the project. JB, EC and JS performed simulation. MG and JS trained and evaluated the models, and JS performed the hardware optimisation. SP supervised the work.

\section*{Declarations}


{\bf Competing Interests} The authors have no competing interests to declare.

\bibliography{sn-bibliography}

\end{document}